\documentclass{gji2}
\usepackage{timet}
\usepackage{graphicx}
\usepackage{amsmath}
\usepackage{amssymb}
\usepackage{flushend}

\title[Early Earthquake Detection with a Dual Torsion-Beam Gravimeter]
  {Early Earthquake Detection with a Dual Torsion-Beam Gravimeter}
\author[D.J. McManus]
  {D. J. McManus, P. W. F. Forsyth, N. A. Holland, R. L. Ward, D. A. Shaddock, 
  	\\ {\LARGE \textup{D. E. McClelland, B. J. J. Slagmolen}}\\
   Centre of Excellence for Gravitational Wave Discovery, Department of Quantum Science, Australian National
    University, Canberra ACT \emph{2601}, Australia.\\Email: david.mcmanus@anu.edu.au; bram.slagmolen@anu.edu.au}
\date{}
\pagerange{\pageref{firstpage}--\pageref{lastpage}}
\volume{}
\pubyear{2018}

\begin{document}

\label{firstpage}

\maketitle

\begin{summary}
Ground mass is redistributed during an earthquake causing the local gravitational potential to change. These gravitational fluctuations travel at the speed of light meaning they will arrive at a remote location significantly earlier than the fastest seismic waves. If these gravitational signals are measured by a gravimeter then early warning can be provided for an imminent earthquake. Earlier detection of earthquakes could be used to protect crucial infrastructure and save lives. The Torsion Pendulum Dual Oscillator (TorPeDO) is a gravity gradient sensor that has been constructed at the Australian National University. In this article we investigate the feasibility of measuring prompt gravitational transients from earthquakes with the TorPeDO. We simulated the response of the sensor to these signals and inserted these responses into scaled TorPeDO strain data to test their detection using a matched filter search. This simulation allows us to estimate the signal-to-noise ratio and detection time of the sensor to these transient signals, along with the influence of different detection thresholds on range and detection time. This article also proposes a method of earthquake localisation using TorPeDO sensors without the need for accurate signal timing. A real-time estimate of earthquake magnitude can be produced by combining this calculated location with TorPeDO strain data. We find that a TorPeDO system operating at design sensitivity would measure a moment magnitude 7.1 earthquake, 200~km away, reaching a signal-to-noise ratio of 5 at 15.7~s after the event starts. This will provide roughly 50.96~s of warning before the arrival of the first S waves.
\end{summary}


\section{Introduction}

%During an earthquake, tectonic forces along a fault exceed friction and cause a rupture. The resulting elastic energy is then released in the form of seismic waves. 
Earthquakes and the resulting seismic waves cause large-scale damage to buildings and infrastructure, as well as injury and death. Since 2011 there have been more than 100,000 fatalities from earthquakes and related tsunamis, with more than 300 billion USD in damages caused  \cite{data}. Many of these deaths are not from the earthquake itself but rather from resulting hazards such as the collapse of infrastructure, floods or explosions. While some of these hazards are unavoidable, others can be mitigated or avoided completely if sufficient early warning is provided before the arrival of damaging S waves. An imminent earthquake can be predicted by monitoring ground motion, however a warning can only be provided after the arrival of P waves. 

Earthquakes displace large amounts of mass and are known to produce long term changes in the local gravitational field \cite{iman}. Short term gravitational transients are also produced during the process of rupture, which travel at the speed of light \cite{jh}. These signals are significantly faster than seismic waves, which are typically slower than 7~km/s \cite{shearer}. A high precision gravimeter would be able to measure these transient signals, making it possible to detect earthquakes earlier than seismometers. In many instances even a minute of early warning can be used to significantly mitigate potential damage. Automated triggers can shut off crucial systems or put them in a safe operating mode. High speed trains can be stopped and people in dangerous situations can be alerted to take cover. In this way early warning can prevent injury, death, or damage to assets and infrastructure.

 The coupling from seismic fields into gravitational gradient has been discussed in numerous publications. In 2016, Montagner et al. published results from a blind search of superconducting gravimeter and seismometer data for a prompt gravitational signal caused by the 2011 Tohoku magnitude 9.1 earthquake. They found strong evidence for the existence of such a signal in the data \cite{mont}. In 2017, Valle\'e et al. published results showing consistent measurement of gravitational transients from the 2011 Tohoku earthquake in post-processing of seismometer data, which also accounted for gravitationally induced elastic deformation of the earth at the location of the seismometers \cite{val}. These publications demonstrated the feasibility of early earthquake detection and magnitude estimation by measuring transient gravitational signals during rupture, before the arrival of seismic waves. 

TorPeDO (Torsion Pendulum Dual Oscillator) is a low frequency gravitational force sensor that uses two torsion pendulums as test masses \cite{dm2,dm1}. Since TorPeDO senses changes in the gravitational field, it would be sensitive to the transient gravitational changes which occur during a nearby earthquake. A network of TorPeDO sensors could be used to determine the location of an earthquake as well as estimating its magnitude.

\section{TorPeDO}
\begin{figure}
	\centering
	\includegraphics[width=3in]{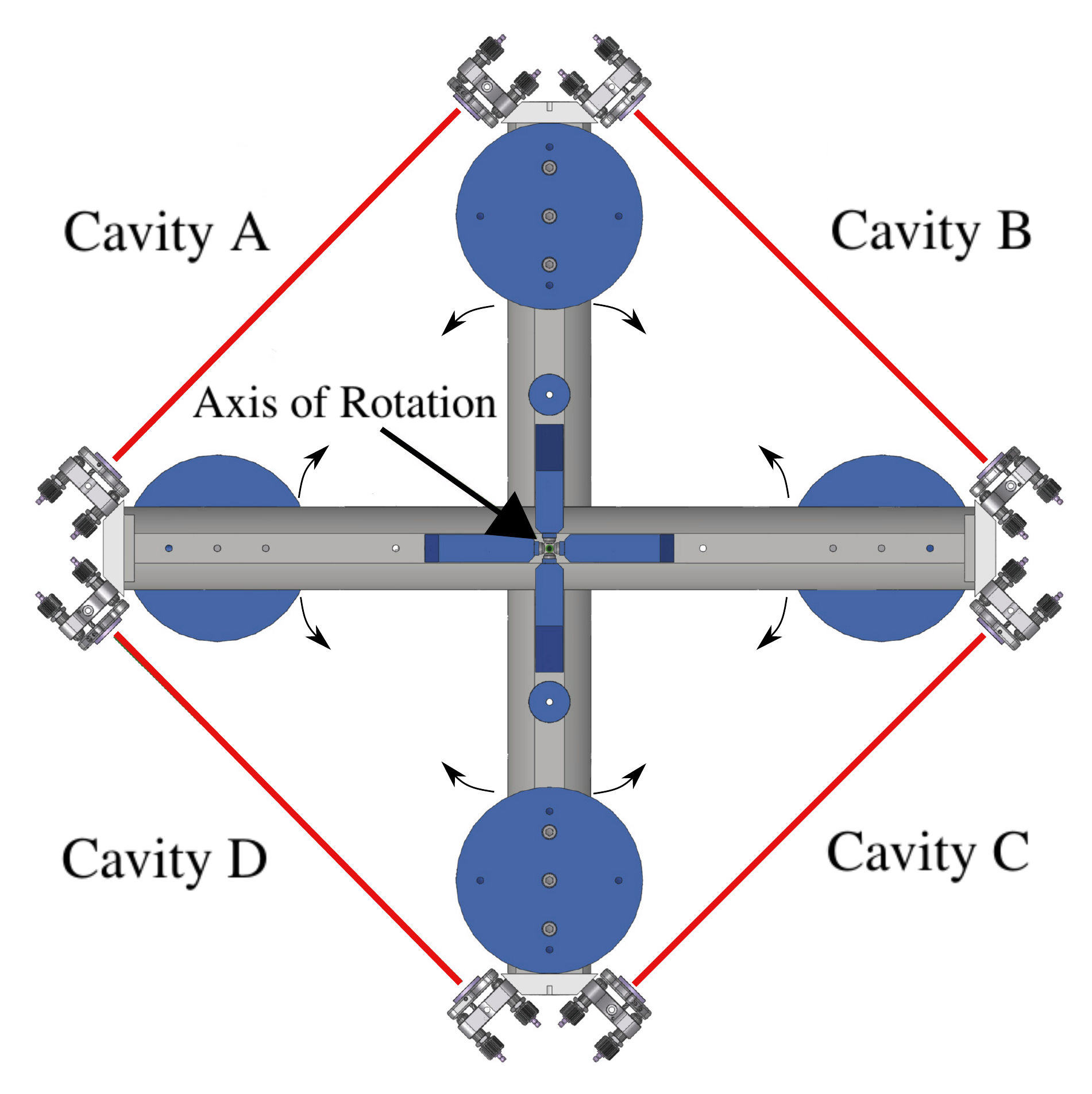}
	\caption{\textsf{Engineering drawing top-view of the TorPeDO showing the two torsion pendulums, their common axis of rotation, and the four Fabry-Perot cavities used for measurement.}}
	\label{fig:cad}
\end{figure}

The TorPeDO design is similar to the TOBA sensor \cite{toba}. The TorPeDO is a gravimeter which utilises two matching torsion pendulums that are suspended orthogonally to each other. These torsion pendulums are free to move under the influence of gravity. A changing gravitational potential causes the two torsion pendulums to rotate differentially. This differential motion is measured optically by monitoring the length change of optical cavities aligned between the ends of the torsion beams. A full description of the sensor design and measurement principle is provided in \cite{dm1}. A top-down engineering drawing of the TorPeDO sensor is shown in Figure \ref{fig:cad} which illustrates the mechanical design. The four optical cavities used for measurement and the axis of rotation are labelled on the figure.

%----------------------- FIGURE Early_warning
%---------------------------

The optical measurement only senses differential motion between the torsion beams, and is insensitive to common motion. Matching the mechanical properties of these two pendulums suppresses the coupling of suspension point motion into the TorPeDO measurement through common mode noise rejection. The two torsion beams of the TorPeDO share a common centre-of-mass position, axis of rotation, and resonant frequency. There are a set of tuning masses installed on both torsion beams which allow for their centre of mass position and resonance frequency to be tuned. These masses can be used to maximise the mechanical common mode rejection.

\section{Prompt Gravity Perturbations from Earthquakes} \label{S:math}
The seismic moment, $M_0$, of an earthquake event is given by Equation \ref{eq:seimom} in the units of Newton meters \cite{mm}.
\begin{equation} 
M_0=\mu D A
\label{eq:seimom}
\end{equation}

Where $\mu$ is the shear modulus, D is the slip displacement and A is the rupture surface area.

A commonly used measure of the scale of an earthquake is the moment magnitude, which is related to the total energy released during an earthquake. It is approximately defined in terms of the total seismic moment as shown in Equation \ref{eq:M_w} \cite{mm}.

\begin{equation}
M_w \approx \frac{2}{3}\big(\log M_0 -9.1\big) 
\label{eq:M_w}
\end{equation}

Earthquakes are difficult to fully characterise because of the complexity of their underlying physical processes. A useful function to characterise the rate of change and severity of an earthquake over time is the source time function, or moment rate function. This is a function of time with units of Nm/s.

\begin{equation} 
M_s[t]=\frac{dM_0}{dt}
\end{equation}

We aim to simulate the gravitational influence of an earthquake on the TorPeDO sensor before the arrival of seismic waves. To model the transient change in potential that occurs during rupture we follow the treatment by J. Harms \cite{jh} for Equations \ref{eq:tensor} to \ref{eq:RMS}. 

First we define the gravity gradient tensor $\textbf{D}(r_0,t)$, which is a tensor defining the rate of change of the gravitational force vector in each direction for a location defined by vector $r_0$ and a given time $t$. This can be defined in terms of the Newtonian gravitational potential $\psi(r_0,t)$ as shown in Equation \ref{eq:tensor}, where $\otimes$ is the Kronecker product or tensor product.

\begin{equation}
\textbf{D}(r_0,t)=-(\nabla \otimes \nabla) \delta \psi(r_0,t)
\label{eq:tensor}
\end{equation}

For the time-frame before the arrival of seismic waves, all parts of this tensor disappear except for one, leading to the approximation shown in Equation \ref{eq:tensor_red}.

\begin{equation}
\textbf{D}(r_0,t)\approx \frac{6G}{r_0^5}S(\theta,\phi)\int_{0}^{t}du\thinspace uM_0(t-u)
\label{eq:tensor_red}
\end{equation}

This gravitational change has an angular dependent magnitude. This information is encoded in the function $S(\theta,\phi)$ which is defined in Equation \ref{eq:S_angfunc}. Here $\vec{e}_x$ is a unit vector corresponding to the earthquake fault normal, and $\vec{e}_z$ is a unit vector in the slip direction. $\vec{e}_r$ is the unit vector from the earthquake centre to the sensor.
\begin{equation}
\begin{aligned}
S(\theta,\phi)=5(\vec{e}_x \cdot \vec{e}_r)(\vec{e}_z \cdot \vec{e}_r)(3 -7\vec{e}_r \otimes \vec{e}_r) \\+4(\vec{e}_x \otimes \vec{e}_z)_{sym}
+5((\vec{e}_x \times \vec{e}_r)\otimes (\vec{e}_z \times \vec{e}_r))_{sym}
\label{eq:S_angfunc}
\end{aligned}
\end{equation}

Where for any vectors $\vec{a}$ and $\vec{b}$:
\begin{equation}
\begin{aligned}
(\vec{a}\otimes \vec{b})_{sym}=\vec{a}\otimes \vec{b}+\vec{b}\otimes \vec{a}
\end{aligned}
\end{equation}

 To estimate the response of the TorPeDO we calculate the gravitational tidal force, which is given by the second time integral of the gravity gradient tensor. This is because the sensor measures gravitational strain which is the change in distance between free falling test masses in space. This is given by the gravity-strain tensor, shown in Equation \ref{eq:strain_tensor}. 

\begin{equation}
\textbf{h} (r_0,t)=\int\int\textbf{D}(r_0,t)~dt^2
\label{eq:strain_tensor}
\end{equation}
Another way to express the strain measured by the sensor before the arrival of seismic waves is to rewrite this result using Equation \ref{eq:tensor_red}, as shown in Equation \ref{eq:sensor_acc}.
\begin{equation}
\ddot{\textbf{h}} (r_0,t)=\frac{6G}{r_0^5}S(\theta,\phi)\int_{0}^{t}du\thinspace uM_0(t-u)
\label{eq:sensor_acc}
\end{equation}

 The measured sensor strain is found by taking the gravitational acceleration from Equation \ref{eq:sensor_acc} and transforming it using unit vectors $e_1$ and $e_2$ which are aligned with the direction of the two torsion pendulums. This is shown in Equation \ref{eq:trans1}, with $e_1^r$ and $e_2^r$ representing those vectors rotated by $90^{\circ}$.
\begin{equation}
\textbf{h}_{\times}(r_0,t)=(e_1^T \cdot \textbf{h}(r_0,t)\cdot e_1^r - e_2^T \cdot \textbf{h}(r_0,t)\cdot e_2^r )/2 
\label{eq:trans1}
\end{equation}
In the case of the TorPeDO these vectors are orthogonal to each other. So the transformation can also be expressed as in Equation \ref{eq:trans}.

\begin{equation}
\textbf{h}_{\times}(r_0,t)=(e_1^T \cdot \textbf{h}(r_0,t)\cdot e_2 + e_2^T \cdot \textbf{h}(r_0,t)\cdot e_1 )/2 
\label{eq:trans}
\end{equation}

The orientation of the TorPeDO sensor influences the result of these equations. The magnitude of cavity length change is dependent on the angle between the arm cavities and the gravitational gradient. A standard rotation transformation of the form $R^T \textbf{h}(r_0,t)R$ can be applied in Equations \ref{eq:trans1} and \ref{eq:trans} to get different sensor orientations. The angular dependent sensitivity of the TorPeDO is a quadrupole pattern of the shape $h\propto \sin(2\theta)$ for in-plane gradients.

For simplified magnitude estimates where we may not have an exact orientation in mind for the sensor or the slip direction, we can use Equation \ref{eq:RMS} which gives the RMS amplitude strain over the different detector and fault orientations for a given source function at time $t$.  
\begin{equation}
h(t)=\frac{6\sqrt{14/5}G}{r^5}I_4\big[M_0(t)\big]
\label{eq:RMS}
\end{equation}

The estimated strain from Equation \ref{eq:RMS} was filtered through a transfer function with the mechanical parameters of the TorPeDO system to simulate the mechanical response of the sensor. 
 %Equation \ref{eq:mech} shows the transfer function of a generic mechanical oscillator, where $\omega_0$ is the resonance frequency, K is a gain value, $\gamma$ is the damping factor, and $s$ is the complex frequency variable $s=\sigma +i\omega $. 

%\begin{equation}
%H(s)=\frac{K~\omega_0^2}{s^2+2\gamma\omega_0~s+\omega_0}
%\label{eq:mech}
%\end{equation}

\section{Signal Detection} \label{S:detect}
Detection of prompt gravitational signals from earthquakes first requires improving the sensitivity of the TorPeDO to $10^{-14}~\text{Hz}^{-1/2}$ in the region from $0.1-1~\text{Hz}$. This level of sensitivity requires actively mitigating the influence of Newtonian noise on the sensor \cite{LF}. Newtonian noise is caused by ambient gravity gradient changes close to the sensor.

%----------------------------------------- FIGURE Time Series ----------------------------
\begin{figure*}
	\begin{center}
		\includegraphics[width=1.02\textwidth]{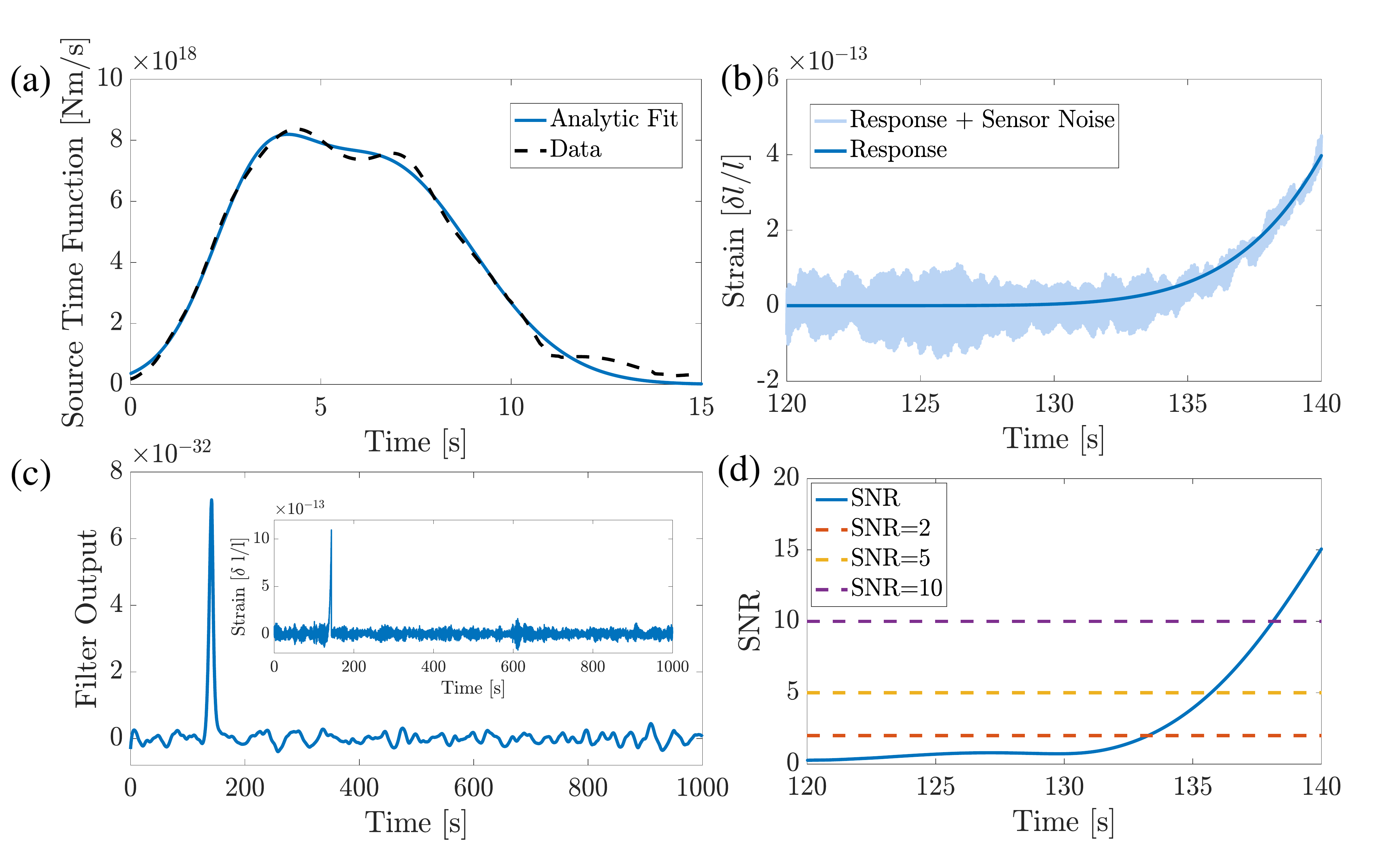}
		\caption[Time series response of the TorPeDO]{Simulated TorPeDO response and detection of a $M_{w}=7.1$ earthquake located 200~km away. \textbf{Figure \ref{fig:TS} (a)} shows the source time function of the earthquake used as an input.  The blue curve is an analytic fit to actual earthquake data from \cite{Val2} shown in black for a 15~km deep, $M_{w}=7.1$ earthquake. \textbf{Figure \ref{fig:TS} (b)} shows the modelled strain response of the TorPeDO to the earthquake from (a). The dark blue line is the modelled response and the light blue plot shows the response inserted into scaled noise from the TorPeDO. The earthquake starts at t=120~s in the noise time-series. Only the transient signal ($t<r_0/\alpha$) is inserted for this simulation. \textbf{Figure \ref{fig:TS} (c)} is a plot of the matched filter output from a search of the noisy time-series data shown in the plot insert. In this simulation the template used is a replica of the signal shown in (b). In \textbf{Figure \ref{fig:TS} (d)} the SNR of the matched filter output from (c) is plotted over time. The dashed lines indicate different SNR thresholds.}
		\label{fig:TS}
	\end{center}
\end{figure*}
%----------------------------------------- FIGURE Time Series----------------------------
The detection method and trigger conditions used for an earthquake alert will influence the warning time and the detection range. In the following analysis we estimate the response of the TorPeDO to gravitational signals from earthquakes of varying magnitudes and distances using the method described in Section \ref{S:math}. Figure \ref{fig:TS} shows the modelled sensor response to a nearby earthquake. The source time function of the event is shown in Figure \ref{fig:TS} (a). The data is from a $M_w=7.1$ earthquake that occurred on 23/10/2011 at latitude 38.72, longitude 43.51, at a depth of 15~km. This data was taken from the SCARDEC Source Time Functions Database \cite{Val2}. The analytic function is a least squares fit of the sum of two skewed Gaussian functions to this data.  Figure \ref{fig:TS} (b) shows the modelled TorPeDO response to this earthquake at a distance of 200~km. The response continues to grow after there is no further change in seismic moment. This is because the seismic waves from the event get closer to the detector, increasing their gravitational influence over time. 

 Once generated these signals can be extracted using the technique of matched filtering in order to estimate the signal-to-noise ratio (SNR), detection range, and trigger time of an early earthquake alert. Matched filtering is a powerful technique for extracting signals of known characteristics from low SNR data. The LIGO collaboration has successfully used matched filtering techniques to extract gravitational wave signals from their data, and estimate their corresponding source parameters \cite{GW1} \cite{GW2}. Matched filtering involves checking the correlation between a known signal template and output data. This is done by convolving the sensor output with a conjugated time-reversed version of the desired signal template. This process will ideally return a low output when the template is compared to noise, but will produce a high correlation peak if the template matches with a signal of the same form in the data. The matched filter output is given by $y$ in Equation \ref{eq:MF} \cite{turin}. Here $q$ is the filter template and $x$ is the data.
 
\begin{equation}
y(t)=\int_{-\infty}^{\infty}q(\tau)x(t-\tau)d\tau
\label{eq:MF}
\end{equation}

%\begin{center}
%	\begin{table}[h]
%		\caption{TorPeDO Mechanical Parameters} \label{tab:params}	
%		\centering
%		\begin{tabular}{@{}c*{50}{c}}
%			Parameter&Value\\
%			\hline
%			K&N/A\\
%		$\omega_0$&0.033~Hz\\
%			 $\gamma$& 1\\
%			\hline
%		\end{tabular}
%	\end{table}
	
%\end{center}

To simulate detector noise, TorPeDO prototype strain data was taken and scaled to design sensitivity levels ($10^{-14}~\text{Hz}^{-1/2}$ at 1~Hz). This noise time series was used to provide an estimate of the SNR and detection time that we should expect for these signals. It is worth noting that the strain sensitivity shape and the detection statistics for the TorPeDO may differ at design sensitivity. The TorPeDO earthquake response from Figure \ref{fig:TS} (b) was inserted into this noise, and then a matched filter search was performed using a copy of the earthquake signal as a template. The matched filter output is shown in Figure \ref{fig:TS} (c). The time in the data where the earthquake occurs can clearly be seen in the plot by a tall correlation peak. For this simulation only the transient signal ($t<r_0/\alpha$) was inserted, meaning that the time series returns to sensor noise after the arrival of P waves. In a real earthquake the sensor would be influenced by earthquake related gradient changes for significantly longer. This however does not affect the signal detection time and significance estimate, which depends only on the rate and size of the initial prompt signal compared to typical noise levels. The evolution of the template SNR is plotted over time in Figure \ref{fig:TS} (d). Just like the sensor response, this increases until just before the arrival of P waves.

Table \ref{tab:detection} lists the detection time for different detection thresholds used for the simulation shown in Figure \ref{fig:TS}. The detection time is defined as the time taken for the matched filter SNR to rise above the chosen detection threshold after the start of the earthquake, assuming that the template search is being performed on live data. The detection time will depend on the shape and size of the earthquake, as well as how close the sensor is to the event. Computation time and other system delays are not considered for this analysis.

%----------------------- FIGURE Early_warning
%\begin{figure}
%	\centering
%	\includegraphics[width=3.1in]{Matched_Filter.pdf}
%	\caption{\textsf{Matched filter result of a magnitude 7.1 earthquake 300~km away inserted into simulated noise shaped by the TorPeDO design sensitivity noise budget. In this simulation the template is an exact match for the embedded fake earthquake signal. The horizontal lines indicate the statistical $3~\sigma$ and $5~\sigma$ levels for the template matching against noise with the earthquake signal removed.}}
%	\label{fig:MF}
%\end{figure}
%---------------------------

When choosing the detection threshold, there is a trade-off between minimising the false alarm rate and obtaining the best detection time and range. Regardless of the signal template used, increasing the detection threshold will increase the detection time and reduce the false alarm rate and vice versa. The optimal trigger threshold is subjective and dependent on the application of the warning system. The false-alarm rates have not been calculated as the detector statistics at design sensitivity are unknown.\\

 The detected SNR value is calculated using Equation \ref{eq:SNR}. This value is an amplitude ratio of how well the template matches with the actual signal compared to background noise. 
\begin{equation}
SNR=\frac{|y_s|}{E\{|y_n|\}}
\label{eq:SNR}
\end{equation}
Where $y_s$ is the output value of the template matching with the signal, and $E\{|y_v|\}$ is the expected value of the template matching with only noise and no signal present. This can be re-written in terms of the filter template, $q$. 
\begin{equation}
SNR=\frac{|q^Hs|}{E\{|q^Hn|\}}
\label{eq:SNR2}
\end{equation}

\begin{figure*}
	\begin{center}
		\includegraphics[width=1\textwidth]{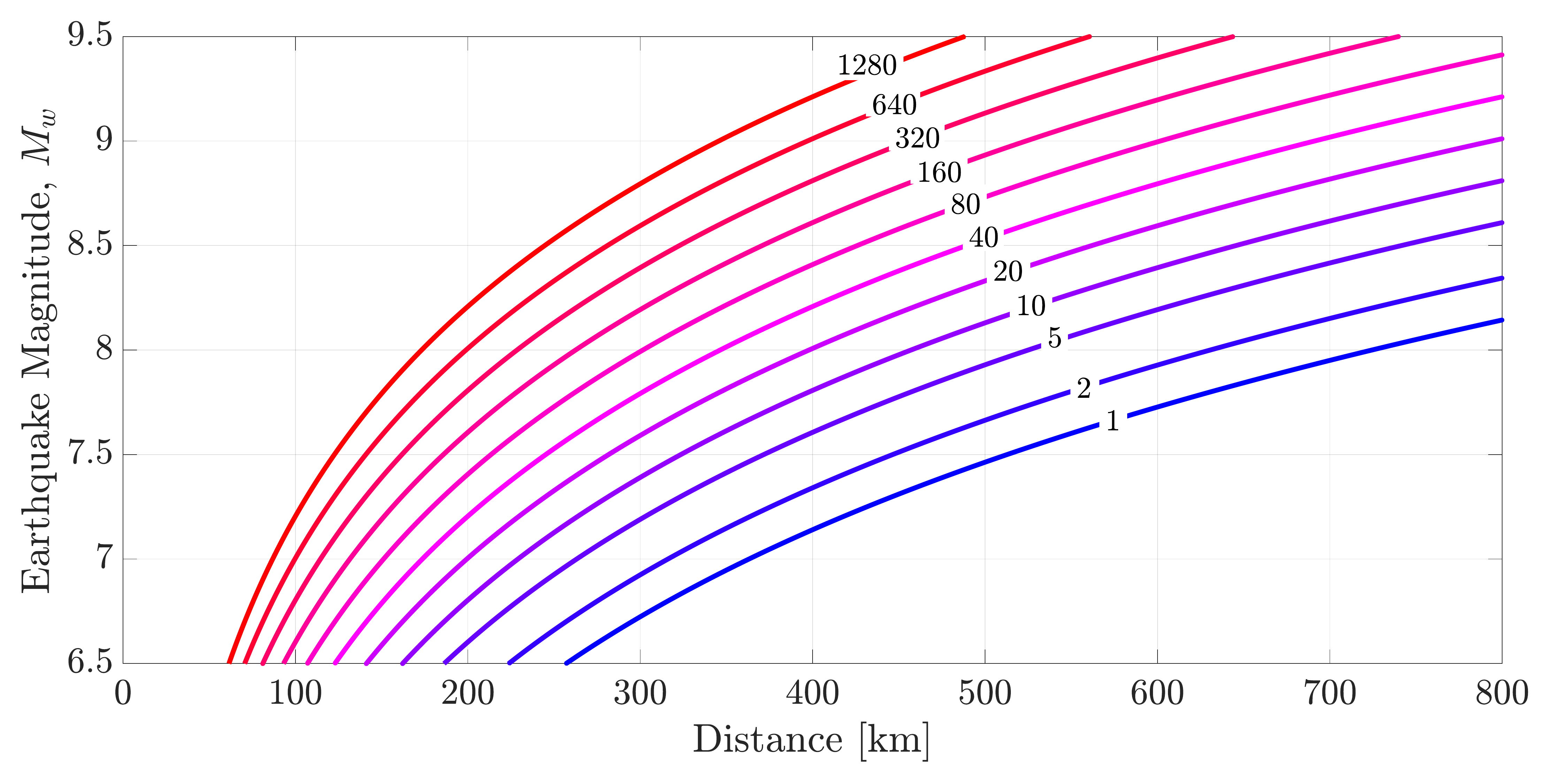}
		\caption[Michelson]{Estimated SNR recorded by the TorPeDO sensor after 15~s for earthquakes of varying distances and magnitudes. Each curve plots a constant SNR, with the SNR value shown as a number along the curve. All plotted values use the same earthquake template from Figure \ref{fig:TS} (a), scaled to different values of $M_w$.}
		\label{fig:SNR}
	\end{center}
\end{figure*}

Where $q^H$ is the conjugate transpose of $q$. Figure \ref{fig:SNR} shows the calculated SNR for different magnitude earthquakes at a given distance from the TorPeDO sensor. For this plot the same earthquake template from Figure \ref{fig:TS} (a) is used and scaled to different values of $M_w$. Higher magnitude earthquakes are likely to have a different shape and time-frame which will influence these values. \\
 %\begin{center}
 	%\begin{table}
 	%	\caption{Signal detection thresholds for the simulation from Figure \ref{fig:MF}. Increasing the detection threshold also increases the time taken to detect an earthquake, but will reduce the chances of a false alarm. These parameters will change for different earthquake profiles. False alarm rates are given in expected number of false alarms per hour.} \label{tab:detection}	
 	%	\centering
 	%	\begin{tabular}{@{}c*{50}{c}}
 	%		Template SNR&Detection Time&False Alarm Rate\\
 	%		\hline
 	%		2& 6.9~s&$0.36~\text{hr}^{-1}$\\
 	%		5&10.7~s&$0.011~\text{hr}^{-1}$\\
 	%		10&14.9~s&$1.22\times10^{-4}~\text{hr}^{-1}$\\
 	%		\hline
 	%	\end{tabular}
 	%\end{table}
 	
\begin{figure*}
	\begin{center}
		\includegraphics[width=1\textwidth]{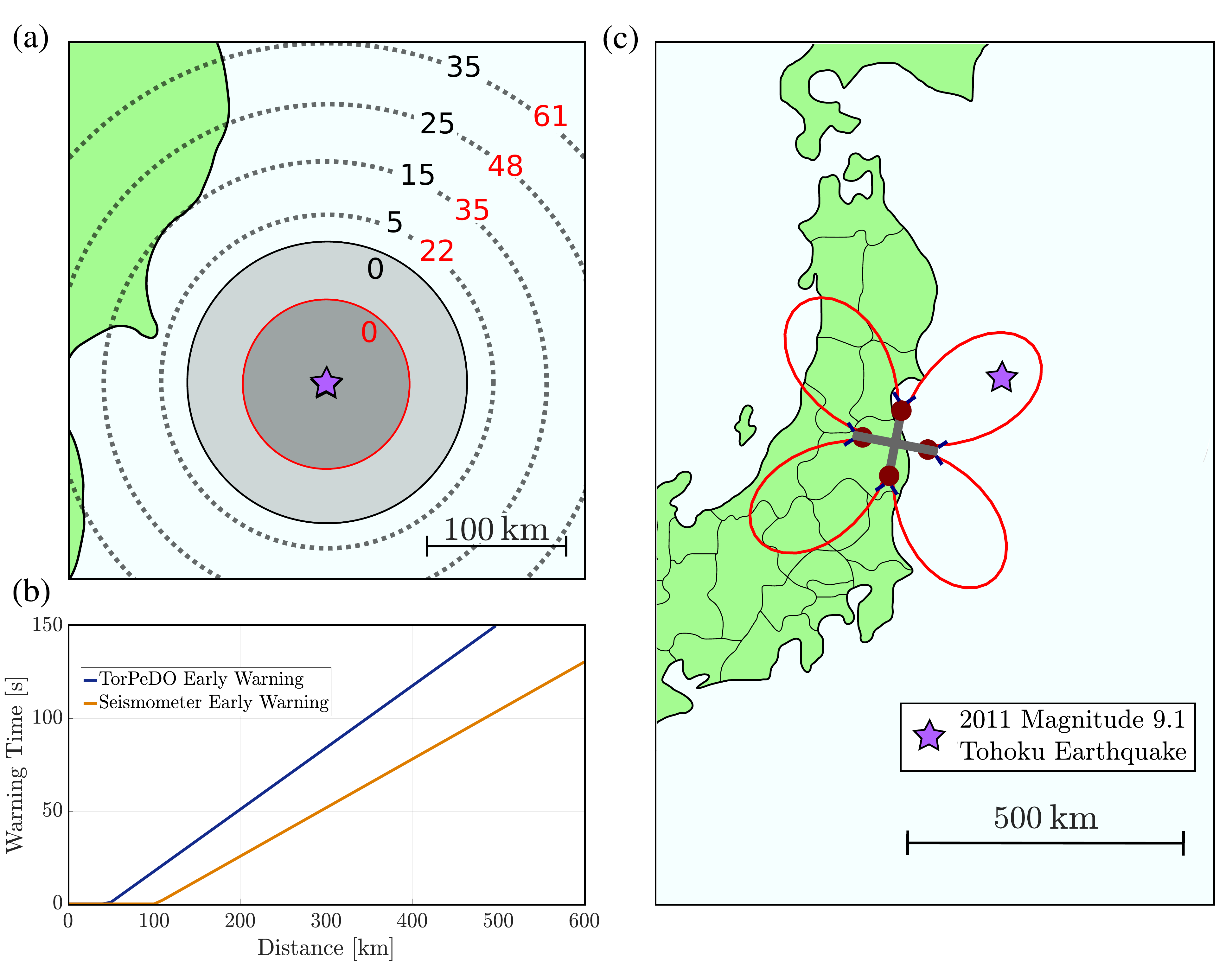}
		\caption[Michelson]{(Colour online) Early warning time and range of the TorPeDO sensor to earthquakes. \textbf{Figure \ref{fig:map} (a)} plots time rings comparing the early warning provided by a gravimeter such as the TorPeDO to that provided by seismometer arrays \cite{JMA}. For each ring the red number indicates the warning time in seconds provided by the TorPeDO before the arrival of S waves. The black number indicates the warning time for seismometer arrays. The red circle labelled zero shows the 'dead-zone' / zero warning region for a TorPeDO warning system, which is the distance that the S waves travel during the time it takes to detect a signal. The black circle is the dead-zone / zero warning region for a seismometer array warning system. \textbf{Figure \ref{fig:map} (b)} shows the same warning time comparison in a plot format. \textbf{Figure \ref{fig:map} (c)} shows the maximum range that the TorPeDO could measure a $M_w=7.1$ earthquake with an SNR of 5, 15~s after the beginning of the earthquake. The sensitivity is plotted as a function of angle on a to-scale map of Japan, with the TorPeDO positioned on the Fukushima coastline. The location of the 2011 magnitude 9.1 Tohoku earthquake is also plotted as a visualisation.}
		\label{fig:map}
	\end{center}
\end{figure*}

 %\end{center}
 \begin{center}
  \begin{table}
  	\caption{SNR detection thresholds for the simulation from Figure \ref{fig:TS}. Increasing the detection threshold also increases the time taken to detect an earthquake, but will reduce the chances of a false alarm. These parameters will change for different earthquake profiles. The detection time is calculated from the start of the earthquake to the point where the measured SNR breaks the listed threshold.} \label{tab:detection}	
  	\centering
  	\begin{tabular}{@{}c*{50}{c}}
  		Template SNR&Detection Time\\
  		\hline
  		2& 13.3~s\\
  		5&15.7~s\\
  		10&18.1~s\\
  		\hline
  	\end{tabular}
  \end{table}
  
  \end{center}
The range of the sensor is influenced by the statistical threshold used for detection, and depends on the magnitude of signals that must be measured. The estimated range of the TorPeDO sensor to a moment magnitude 7.1 earthquake is shown on a map of Japan in Figure \ref{fig:map} (c). The range here is defined as the maximum distance where an SNR of 5 will be recorded after 15~s. The maximum range in the figure is roughly 303~km. The sensor is positioned on the Fukushima coast with the 2011 Tohoku magnitude 9.1 earthquake plotted as an example event.  This figure shows how the TorPeDO sensitivity changes with torsion bar alignment as discussed in Section \ref{S:math}. The sensitivity is maximised when the gravitational gradient angle is at $45^{\circ}$ to both torsion beams. The device is insensitive to sources located along the axis of either beam. This is because gravitational attraction or repulsion along this axis will cause both beams to translate in common mode but not rotate differentially. 

%----------------------------------------- FIGURE Michelson Interferometer ----------------------------
%----------------------------------------- FIGURE Michelson Interferometer ----------------------------

%----------------------------------------- FIGURE EQ Map  ----------------------------
%----------------------------------------- FIGURE EQ Map ----------------------------

\section{Early Warning}
The warning time achieved by measuring these gravitational signals is straightforward to calculate. The time advantage comes from the difference in propagation speed of the gravitational signal and that of the emitted seismic waves. The trigger time of the detector should also be taken into account, which will depend on both the source and the detection trigger as discussed in Section \ref{S:detect}.

\subsection{Signal Travel Time}
We know from GW170817, the detection of gravitational waves from a neutron star binary, that gravitational signals travel at the speed of light \cite{GW3}. Therefore the travel time of the gravitational signal is given by the speed of light, $c$, and the distance between the sensor and earthquake. The first seismic waves to arrive after an earthquake are the P waves. These are pressure waves propagating through the earth and are the fastest travelling seismic waves. The speed of the P waves, $\alpha$, is given by Equation \ref{eq:p_vel} \cite{shearer}.

\begin{equation}
\alpha=\sqrt{\frac{\lambda_{L} +2\mu_{L}}{\rho}}
\label{eq:p_vel}
\end{equation}
Where $\rho$ is the ground density, $\mu_L$ is the shear modulus, and $\lambda_L$ is the Lam\'e parameter of the wave medium.

The speed of the slower and more destructive S waves, $\beta$, is given by Equation \ref{eq:s_vel} \cite{shearer}. 

\begin{equation}
\beta=\sqrt{\frac{\mu_{L}}{\rho}}
\label{eq:s_vel}
\end{equation}

Most earthquake early warning systems use the P-wave signals to provide an alert. Comparing Equation \ref{eq:p_vel} to the speed of light gives the travel-time advantage of the gravitational signal over P waves.
\begin{equation}
t_{adv,~p}=r_0\Big(\frac{1}{\alpha}-\frac{1}{c}\Big)=r_0\Big(\sqrt{\frac{\rho}{\lambda_{L} +2\mu_{L}}} -\frac{1}{c}\Big)
\label{eq:p2}
\end{equation}

Where $r_0$ is the distance of the earthquake to the sensor. The travel time-advantage of the gravitational signal over the first damaging seismic waves (S waves) is calculated in the same way, this time using Equation \ref{eq:s_vel}.

\begin{equation}
t_{adv,~s}=r_0\Big(\frac{1}{\beta}-\frac{1}{c}\Big)=r_0 \Big(\sqrt{\frac{\rho}{\mu_{L}}} -\frac{1}{c}\Big)
\label{eq:s2}
\end{equation}

Figure \ref{fig:map} (b) compares the warning time difference between a gravimeter early earthquake system to a currently used seismometer early earthquake system in Japan \cite{JMA}. The trigger time for the TorPeDO here is set to 15~s, since this is the detection time used in Figure \ref{fig:SNR} to define the measured SNR. This comparison is shown graphically in Figure \ref{fig:map} (a). The time rings show the TorPeDO warning time in red numbers at different distances from an earthquake, compared to the black numbers indicating the warning time of the seismometer array system. The seismometer warning time is adapted from \cite{JMA} and \cite{yamasaki}.

%----------------------- FIGURE Early_warning
%\begin{figure}
%	\centering
%	\includegraphics[width=3.4in]{EEW.pdf}
%	\caption{\textsf{Comparison in warning time for a potential TorPeDO early warning system and a current seismometer early warning system used in Japan [citation]}}
%	\label{fig:times}
%\end{figure}
%---------------------------

%----------------------- FIGURE Early_warning
%\begin{figure}
%	\centering
%	\includegraphics[width=3.2in]{EEQ_comp_3.pdf}
%	\caption{\textsf{Comparison between TorPeDO earthquake early warning and Japan's seismometer based early earthquake system. The numbered red rings show predicted TorPeDO early warning provided at this distance in seconds. The black rings show the early warning of the seismometer based system. [I don't have copyright to this image, it will need to be changed]}}
%	\label{fig:rings}
%\end{figure}
%---------------------------

\section{Localisation \& Magnitude Estimation} \label{S:loc}
Estimating the location and magnitude of imminent earthquakes is a crucial aspect of providing early warning. The appropriate response to an oncoming earthquake may depend on the size of the event, and also how much time there is before the arrival of damaging seismic waves. An accurate estimate of earthquake magnitude requires knowledge of its location. This is because gravitational signals from a closer, smaller earthquake can be confused with those from a larger earthquake located further away.

%----------------------------------------- FIGURE EQ Map  ----------------------------
\begin{figure*}
	\begin{center}
		\includegraphics[width=1\textwidth]{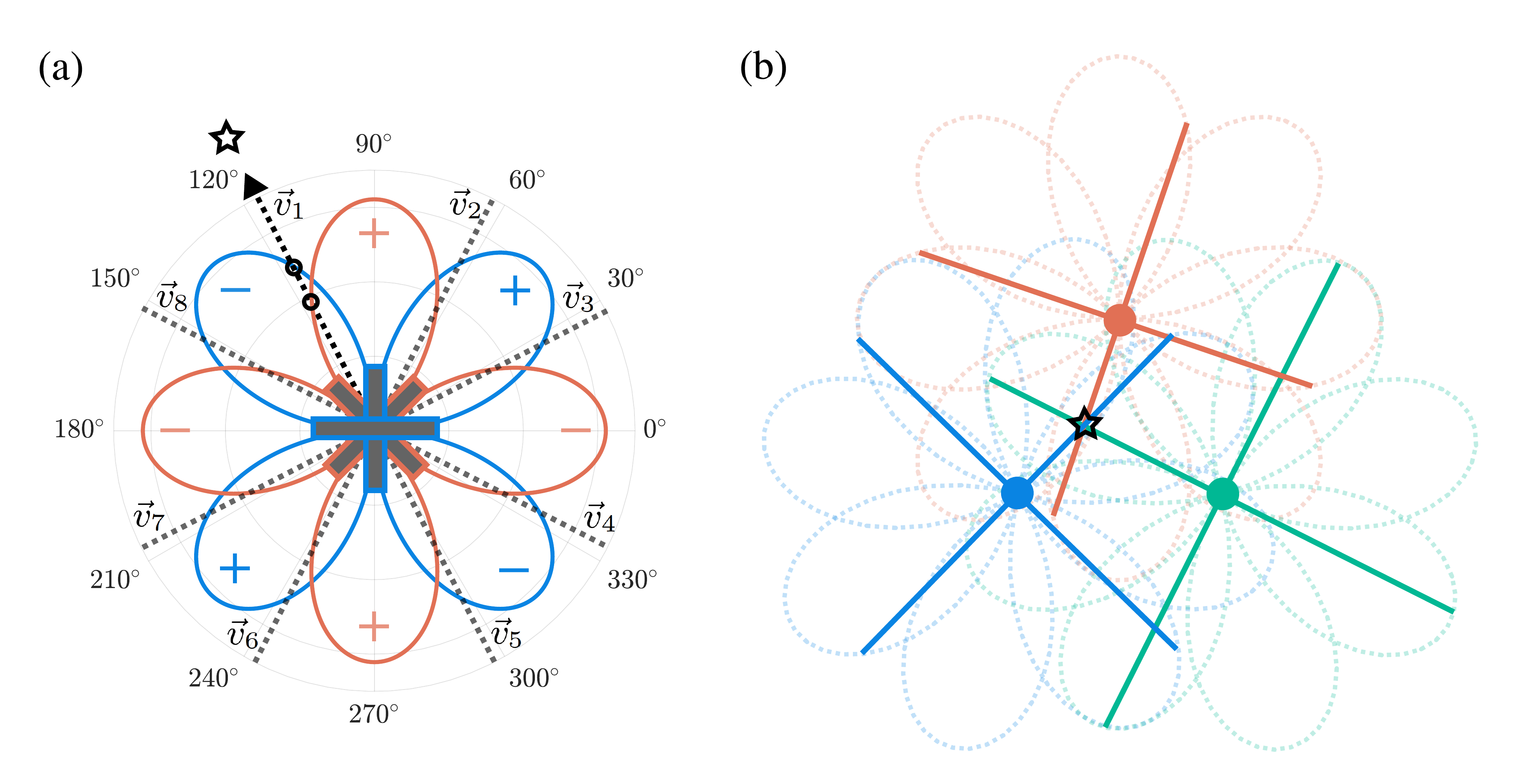}
		\caption[Localisation]{(Colour online) Earthquake localisation using relative magnitudes recorded by co-located sensors offset by 45$^{\circ}$. \textbf{Figure \ref{fig:local} (a)} shows a station with two TorPeDOs (shown as red and blue), with both sensors recording a signal from an event in the direction $\vec{v}_1$. The relative signal magnitude recorded by each TorPeDO is indicated by the points where $\vec{v}_1$ intersects the sensor's plotted angular sensitivity. Each recorded ratio of magnitudes corresponds to 8 possible angles (or 4 possible angles if the event is along a bar axis), shown visually in this example by vectors $\vec{v}_1,...,\vec{v}_8$. Using the rotation polarity of each sensor narrows this down to 4 possible angles, each at $90^{\circ}$ from each other. For the measurement in \ref{fig:local} (a), the possible solutions are $\vec{v}_1, \thinspace \vec{v}_3,\thinspace\vec{v}_5,\thinspace \& \thinspace \vec{v}_7$. \textbf{Figure \ref{fig:local} (b)} shows how multiple stations, each with a pair of sensors, can be used to uniquely find the earthquake location. 3 stations shown as blue, red, and green circles are located in an equilateral triangle of side length 250~km. The star indicates the location of an earthquake, with each of the three stations recording a measurement. For each sensor pair there are 4 possible angles the earthquake signal could come from, however by combining the information from each pair we get a unique solution for the location of the earthquake. There are three other locations in this example where the vectors from two stations intersect, but a third sensor measurement (or lack of measurement) uniquely locates the event.}
		\label{fig:local}
	\end{center}
\end{figure*}
%----------------------------------------- FIGURE EQ Map ----------------------------

Localisation of earthquake sources cannot be reliably done with a single sensor. As plotted in Figure \ref{fig:map}, the sensor is not uniformly sensitive at all angles. This means that even if you know the magnitude of an earthquake source, attenuation due to distance cannot be distinguished from attenuation due to angle. For this reason even an estimate of distance requires multiple sensors.

Localisation with multiple sensors can usually be achieved using triangulation of recorded signal time delays. In the case of TorPeDO sensor arrays, time delay measurements are extremely difficult because of the low signal frequency. In the absence of any measured cycles or turning points, the relative phase of each measurement can only be estimated using the relative magnitude between each sensor over time. This magnitude difference will be extremely small and is influenced by the orientation of the source and the sensor, which is unknown. For these reasons it is unrealistic to expect accurate event localisation from signal timing triangulation.

\subsection{Localisation Using Relative Signal Amplitude} \label{S:loc_tech}
 Here we propose an alternative method for earthquake localisation using multiple sensors. Suppose that two TorPeDO sensors are positioned at the same location, but have an angular offset of $45^{\circ}$ with respect to each other. In this configuration, the relative signal amplitude recorded by each TorPeDO will indicate the possible directions of an oncoming signal due to the angular sensitivity of the two sensors. Figure \ref{fig:local} (a) shows this concept visually, with any signal detected by both sensors corresponding to 8 different possible directions. The polarity of each sensor measurement can be used to distinguish between some of these solutions. Every combination of signal polarity has only 2 vector solutions. An attractive potential in one direction is indistinguishable from a repulsive potential in a direction with the opposite polarity.  This means that any measured signal can only be narrowed down to 4 possible angles, with each solution spaced by $90^{\circ}$. 

Using multiple stations, each consisting of two co-located sensors, it is possible to obtain a unique solution for the earthquake location. This is done by solving for a unique set of coordinates that satisfies the measured information at each station. In Figure \ref{fig:local} (b) we can see three stations positioned in an equilateral triangle of side length 250~km. An earthquake is measured by all three stations and the relative signal amplitude at each station points to a unique location, indicated by the star, where the event must have occurred. 

It is important to note that this technique solves for the projection of the earthquake location onto the plane of the sensors, it doesn't tell you the depth. The depth can be calculated from the relative signal strengths between different stations once the corresponding surface location is known.\\

 Suppose that $v^n_i$ is any solution vector $\vec{v}_i$ from Figure \ref{fig:local} (a) of station number $n$. The possible locations of the earthquake are given by the points, $P$, where a solution vector from each detecting station intersects.
 \begin{equation}
 P=\{v^1_i=v^2_j=...=v^n_k\}
 \label{eq:points}
 \end{equation}
 These possible solutions are narrowed down by looking only at areas within the range of all detecting stations, and ruling out regions within the range of non-detecting stations. If we define $M_i$ as the sensing range of any station $i$ which records a measurement, and $N_j$ as the sensing range of any station $j$ which does not record a measurement, then this area is defined as $A$ in equation \ref{eq:region}.  
 \begin{equation}
A=\bigcap_i M_i \medspace/\medspace \bigcup_j N_j
\label{eq:region}
 \end{equation}
 Therefore the solutions for the earthquake location from Equation \ref{eq:points} in this range are given by Equation \ref{eq:L}. 
\begin{equation}
L=P\cap A
\label{eq:L}
\end{equation}

It is possible to position sensors so that any measurement made by at least two stations results in a unique solution for the earthquake location. The configuration shown in Figure \ref{fig:local} (b) results in a triangular detecting area of roughly $1.56\times10^5~\text{km}^2$ with each station located near the mid-point of a $600~\text{km}$ long edge. Often a measurement made by even a single station on the outskirts of an array can be localised down to a unique angle and bounded by the sensor range. This is because the other possible angles for the measurement may be covered by the sensing area of other stations.

\subsection{Estimates of Earthquake Magnitude}
If matched filtering is used to provide an early alert, the magnitude of the event can be estimated by taking the magnitude of the template which matches best with the measured signal. If the event is located using the technique described in Section \ref{S:loc_tech}, then the signal attenuation caused by distance and orientation is known and can be accounted for in the magnitude calculation.

It is likely that a given event may trigger multiple templates above their respective detection threshold. The relative likelihoods of the triggered templates should inform an appropriate response to the alert. A lower bound for the earthquake magnitude is provided by the smallest triggered template. As time increases some templates will be ruled out, and others become more likely. Estimates of earthquake magnitude will therefore become more accurate as the earthquake evolves

\subsection{Uncertainties in Location and Magnitude Estimates}
The method described in Section \ref{S:loc_tech} uses relative signal amplitude to locate an earthquake. Amplitude noise in the sensor will therefore become position uncertainty when the location is calculated. If we assume that the amplitude noise in each sensor is uncorrelated then Equation \ref{eq:ang_error} describes the angular uncertainty, $\Delta \theta$, calculated by each two-sensor station for given strain amplitude uncertainties $\Delta h_1$ and $\Delta h_2$ of the two TorPeDOs at the station.

\begin{equation}
\Delta \theta =\frac{1}{2}\sqrt{\frac{\Delta h_1^2 \tan^2(2\theta^*)}{h_1^{*2}}+\frac{\Delta h_2^2\tan^2(2\theta^*+\frac{\pi}{2})}{h_2^{*2}}}, \thickspace h_1^*,\thinspace h_2^* \not = 0
\label{eq:ang_error}
\end{equation}

Where $\theta^*$ is the measured angle estimate and $h^*_1,\thinspace h^*_2$ are the magnitudes of the signals recorded by the two TorPeDO sensors. 

This calculation is based on a partial derivative linear approximation, and as such the uncertainty function goes to infinity at the angles where the sensitivity is 0. In practice, the angular uncertainty actually has an upper limit. If TorPeDO 1 at the station takes a measurement, then we know that the event is located in the angular region where the corresponding signal for TorPeDO 2 is below noise and vice versa. Therefore instead of tending to infinity in the case where no measurement is made, the angular uncertainty is actually capped out by the size of the segment where the measurement will be below noise. Equation \ref{eq:ang_cap} gives an upper bound for this uncertainty, assuming that a detection is made at TorPeDO 1 and the corresponding signal for TorPeDO 2 is less than $2\Delta h_2$

\begin{equation}
\label{eq:ang_cap}
\Delta \theta_{max} \leq \frac{\Delta h_2}{h_1^*}
\end{equation}

We therefore should expect an angular uncertainty described by Equation \ref{eq:ang_error}, except limited by Equation \ref{eq:ang_cap} in the cases where only 1 sensor records a measurement above noise.

This error in angle propagates into the location calculation. Since the calculated angles for each station have uncertainty, then there is no guarantee that the vectors from all stations will exactly intersect at the location of the earthquake. In this case the point which has the minimum total distance to all detection vectors is used for the earthquake location. Equation \ref{eq:loc_error} describes the error in location, $\Delta r$, caused by sensor noise. Here $r*$ is the distance from the station to the estimated earthquake location, and $\theta*$ is the calculated angle of the event from the station.
\begin{equation}
\Delta r = \frac{E\{r^*\Delta \theta \}}{\sqrt{N}}
\label{eq:loc_error}
\end{equation}

We see that this localisation error scales as $1/\sqrt{N}$ where N is the number of detecting sensors, and so the location is known more accurately as the number of sensors increases.

Suppose the earthquake from Figure \ref{fig:TS} was located in the middle of 3 sensors in a triangular formation with a 250~km side length. In this case we find that the expected angular error, $\Delta \theta$, per station would be 0.0294~radians averaging across all station alignments. The expected location uncertainty, $\Delta r$, in this situation would be a 5.86~km radius around the earthquake location.

The earthquake magnitude estimate will depend on $\Delta M_0$, which is influenced by errors in position and amplitude. Equation \ref{eq:mag_error} gives the uncertainty of the moment magnitude in terms of $M_0$.

\begin{equation}
\Delta M_w=\frac{2\Delta M_0}{M_0^* \log{10^3}}
\label{eq:mag_error}
\end{equation}
Where $M_0^*$ is the estimated value of $M_0$ based on measurements of the earthquake profile. 
%\begin{equation}
%\sigma_x= \frac{c~\sigma_{timing}}{\sqrt{N}} \approx \frac{600~m}{\sqrt{N}}
%\label{eq:timing}
%\end{equation}

%\section{Magnitude Estimation}

%Once the source is located using the method from Section 6, Equation \ref{eq:sensor_acc} and Equation \ref{eq:S_angfunc} define the change in magnitude with distance and angle. This information along with a calibrated sensor signal is sufficient to extract the change in seismic moment over time. This can then be used to provide a lower bound on the moment magnitude of the measured earthquake using Equation \ref{eq:M_w}.

\section{Conclusions}
We have investigated the feasibility of the TorPeDO sensor as part of an early earthquake warning system, and compared it to existing seismometer based systems. Early earthquake warning by detecting prompt gravitational transients offers a significant time advantage over seismometer based methods. A design sensitivity TorPeDO sensor should be able to measure a moment magnitude 7.1 earthquake, 200~km away, reaching a signal-to-noise ratio of 5 at 15.7~s after the event starts. This provides roughly 50.96~s of warning before the arrival of the first S waves. 

Matched filtering was explored as a method of signal detection and extraction. The detection threshold used influences the sensor range and detection time. 

The earthquake location can be determined using a relative amplitude measurement from multiple sets of co-located sensors. Combining the calculated location with the template parameters from each measured signal can be used to provide a live estimate and bound of the earthquake magnitude.

\begin{acknowledgments}
We acknowledge Ayaka Shoda from the National Astronomical Observatory of Japan for providing assistance with the control scheme of the TorPeDO. We also would like to thank Giles Hammond from the University of Glasgow for his assistance with future suspension design of the sensor. We would like to thank Stephen Cox, Jan Harms and Bernard Whiting for helpful correspondence during the preparation of this article. The authors would like to acknowledge support from the Australian Research Council grant FT130100329 and DP160100760. This paper has been assigned LIGO document number LIGO-P1800090.

\end{acknowledgments}

%\appendix
%\section{blank}

\label{lastpage}

\end{document}